# Impact of Band Structure on Wave Function Dissipation in Field Emission Resonance

Wei-Bin Su[1*], Shin-Ming Lu[1], Ho-Hsiang Chang[1], Horng-Tay Jeng[2,1,3*], Wen-Yuan Chan[1], Pei-Cheng Jiang[1], Kung-Hsuan Lin[1], and Chia-Seng Chang[1]

[1]*Institute of Physics, Academia Sinica, Nankang, Taipei 11529, Taiwan*

[2]*Department of Physics, National Tsing Hua University, Hsinchu 30013, Taiwan*

[3]*Physics Division, National Center for Theoretical Sciences, Hsinchu 30013, Taiwan*

We demonstrated on Ag(111) and Ag(100) surfaces that the reciprocal of the field emission resonance (FER) linewidth, which is proportional to the mean lifetime of resonant electrons in FER, may vary with the electric field. The variation on Ag(111) was nearly smooth, whereas that on Ag(100) was sporadic and fluctuated remarkably. This drastic difference can be explained through their dissimilar projected bulk band structures and the ensemble interpretation of quantum mechanics, according to which all resonant electrons are governed by a single wave function. Ag(100) has an energy gap above its vacuum level, whereas Ag(111) does not. Consequently, the dissipation rate of the wave function, which is relevant to the FER linewidth, on Ag(111) was almost stable, whereas that on Ag(100) fluctuated. The fluctuation revealed that the quantum trapping effect and surface dipole layer on Ag(100) surface can be investigated through FER.

Field emission resonance (FER) [1] in scanning tunneling microscopy (STM) [2,3] is a versatile technique for investigating various physical phenomena and properties, such as the atomic structure of an insulator [4], plasmon-assisted electron tunneling [5], resistance at the nanometer scale [6], light emission [7-9], the dynamics and lateral quantization of surface electrons above the vacuum level, the work functions of thin films [10-17], surface reconstructions [18-20], sharpness [8,21,22] and field enhancement factors [23] of STM tips, and band gaps [20]. Recently, we demonstrated on the bulk $MoS_2$ that FER can be used to observe the quantum trapping that leads to the variation in the FER linewidth being as high as one order of magnitude [24]. According to the uncertainty principle, the reciprocal of the FER linewidth is proportional to the mean lifetime of electrons in FER (named resonant electrons). This case on the $MoS_2$ implies that the lifetime can be sensitive to the physical properties of the surface.

In quantum mechanics, wave function (WF) describes the wave behavior of particles. In FER, the WF is a standing wave. Previous studies have demonstrated that resonant electrons can exit FER through surface transmission [20] or light emission [7,8], which implies that the WF is not stationary but can be dissipated. Wave function dissipation (WFD) in FER and the correlation of WFD with surface properties have been seldom investigated. In this Letter, we demonstrated on Ag(111) and Ag(100) surfaces that WFD can be studied by using the FER linewidth and electric field of FER formation. Ag(100) and Ag(111) were selected for comparison because in their projected bulk band structures, Ag(100) has an energy gap above its vacuum level but Ag(111) does not [25]. The results indicated that the dissipation rate of the WF in first-order FER on Ag(111) was higher than that on Ag(100). Furthermore, the dissipation rate on Ag(111) was nearly constant but that on Ag(100) fluctuated remarkably, which indicated that the band structure considerably affected WFD. To explain these results, we adopted the ensemble interpretation of quantum mechanics [26], according to which all resonant electrons in FER are governed by a single WF. In fact, many interpretations of quantum mechanics have been proposed [27]. FER may be a touchstone for verifying various interpretations, similar to the interference pattern in the double-slit experiment [28].

In the experiment, clean Ag(111) and Ag(100) surfaces were prepared using ion sputtering followed by annealing at 600 °C for several cycles, then transferred to ultra-high-vacuum STM operated at 78 or 5.5 K. FERs were observed through *Z-V* spectroscopy by using PtIr tips. The *Z-V* spectrum was differentiated using a numerical method to reveal FERs.

One study demonstrated that FER peaks observed on the reconstructed Au(111) surface exhibits spatial variation [20]. The peak intensity at the ridge region was lower than that at the fcc and hcp regions, but the opposite trend was observed for the intensity of the valley between FER peaks. This intensity difference was attributed to the higher electron transmissivity of the ridge region. In fact, this intensity variation also indicated that the linewidth $\Delta E$ of FERs at the ridge region was greater than that at the fcc and hcp regions. Figure 1(a) shows FER peaks simulated using the Lorentzian function under the same intensity integration, and that for a greater $\Delta E$, the peak intensity is lower but the intensity at the valley between peaks is higher, consistent with the experimental observation on Au(111). Therefore, $\Delta E$ may increase with an increase in transmissivity.

In the ensemble interpretation of quantum mechanics [26], all resonant electrons in FER are considered to be an ensemble governed by a single WF. When the electron transmissivity of the surface is between 0 and 1, because of partial transmission, WFD occurs when the resonant electrons impinge on the surface. Moreover, studies have demonstrated that light can be detected when observing FER [7,8], which implies that the light emission is also a pathway for WFD. A resonant electron eventually leaves FER through transmission or light emission. However, before leaving, the electron moves back and forth in round trips in the STM junction and forms the standing wave necessary to reveal FER. WFD leads to decay of the probability $P(j)$ that resonant electrons remain in FER, where $j$ is the number of the round trips. $P(j)$ can be defined as follows

$$P(j) = (1 - D)^j, \quad (1)$$

where $D$ is the decay rate of the probability per round trip, which is the sum of $\Gamma_t$ (transmissivity) and $\Gamma_l$ (decay rate due to light emission). Equation (1) suggests that the probability that a resonant electron stays in FER for ($j$-1) round trips but leaves FER in the $j^{\text{th}}$ round trip is $P(j-1) - P(j)$. The average number of round trips $< j >$ of all resonant electrons is as follows:

$$< j >= \sum_{j=1}^{j=j_{max}} j[P(j-1) - P(j)], \quad (2)$$

where $j_{max}$ is the maximum number of round trips for which a resonant electron can remain in FER, depending on the number $N$ of all resonant electrons in FER. Figure 1(b) presents a plot of $< j >$ versus $D$, which was obtained using Eqs. (1) and (2) for $N$=$10^3$, $10^4$, and $10^5$. The plots shows a trend that $< j >$ decreases with increasing $D$ and is independent of $N$. This trend can be well fit by a curve representing

$$< j > = 1/D. \quad (3)$$

Because $N$ is approximately $10^{10}$ for FER, the $< j >$ versus $D$ plot in the FER case should follow Eq. (3). Because the ridge region on Au(111) has a higher $\Gamma_t$, $D$ is larger, which results in a smaller $< j >$. A larger $< j >$ indicates a longer mean lifetime, and $< j >$ is linearly proportional to $1/\Delta E$. Therefore, a higher $\Gamma_t$ corresponds to a greater $\Delta E$, which explains the wider FER on the ridge region.

Transmissivity is related to the density of states [20]. Therefore, it can be suggested that $\Gamma_t$ is zero for the electron energy in the energy gap. Because light emission is the only channel for WFD, the $\Delta E$ of FER on a material with an energy gap should be different from that of a material without an energy gap. To verify this, Ag(100) and Ag(111) surfaces were selected because Ag(100) has an energy gap above its vacuum level, whereas Ag(111) does not [25]. Figures 2(a) and 2(b) display typical FER spectra for Ag(111) and Ag(100), respectively, at 78 K under 10 pA. The numbers in the figure denote the order of the FERs. The potential of FERs of order higher than 0 is the external potential typically approximated by a linear potential. Therefore, the energies of FERs are described as follows [29]:

$$E_n = E_{vac} + \alpha F_{FER}^{\frac{2}{3}}(n - \tfrac{1}{4})^{\frac{2}{3}}, \qquad (4)$$

where $E_{vac}$ is the vacuum level, $F_{FER}$ is the electric field of FER formation, and $n$ = 1, 2, 3… is the quantum number equal to the order number, and $\alpha = \left(\frac{\hbar^2}{2m}\right)^{\frac{1}{3}} (\frac{3\pi e}{2})^{\frac{2}{3}}$. Figure 2(c) displays plots of the peak energies of the higher-order FERs in Figs. 2(a) and 2(b) versus $(n - 1/4)^{2/3}$. The results reveal that the data points of two cases can be fit favorably by lines for orders from 1 to 3 but for those beyond order 3, the data deviate obviously from the linear fit. This deviation is attributed to the apex curvature of the STM tip, which enables the formation of FERs (order > 3) under a weaker electric field [22]. The lines in Fig. 2(c) are parallel, indicating that the $F_{FER}$ of forming FERs 1−3 is the same for both spectra. This $F_{FER}$ can also be obtained from the slope of the line. Moreover, in the FER, resonant electrons move back and forth within a distance between the surface and the classical tuning point. This distance $s$ can be calculated as $(E_n - E_{vac})/eF_{FER}$. Consequently, the round-trip time $t$ for resonant electron motion can be calculated using $2s = eF_{FER}\, t^2/m$. By combining this equation with Eq. (4), we obtain the following equation:

$$t = \beta \frac{(n - \frac{1}{4})^{\frac{1}{3}}}{F_{FER}^{\frac{2}{3}}}, \quad (5)$$

where $\beta = (\frac{3m\hbar\pi}{e^2})^{\frac{1}{3}}$. The mean lifetime can be defined as the product of the average number of round trips and the round-trip time ($<j>\ t$), which is proportional to $1/\Delta E$.

Because the energy of FER 1 is in the energy gap of Ag(100), here, we focus on the comparison of FER 1 on Ag(111) and Ag(100). Figures 2(d) and 2(e) display the FER 1 in Figs. 2(a) and 2(b), respectively. Although acquired under the same $F_{FER}$, the FER 1 on Ag(111) was considerably broader than that on Ag(100), as indicated by their $\Delta E$, which was obtained from Lorentzian fittings. $F_{FER}$ can be tuned by adjusting the sharpness of STM tip through the spontaneous change due to the thermal effect [24] (see Supplemental Material [30]) or by applying a voltage pulse. Under the same current, $F_{FER}$ is weaker and the number of FERs is higher when the tip is sharper [22]. Therefore, $\Delta E$ under various $F_{FER}$ can be investigated through FER spectra of various sharpness levels. Figure 3(a) displays that $1/\Delta E$ versus $F_{FER}$ plot, showing that the values on Ag(111) are prominently lower than those on Ag(100) within an $F_{FER}$ range. Moreover, the fluctuation of $1/\Delta E$ on Ag(100) is considerably greater than that on Ag(111). Figure 3(b) displays the $1/\Delta E$ versus $F_{FER}$ plot for Ag(111) solely, which reveals that on average, the data points follows a curve representing $1/\Delta E$, which is proportional to $F_{FER}^{-\frac{2}{3}}$. This proportionality can be confirmed by tuning the current to adjust $F_{FER}$ at 5.5 K under the same tip structure (see Supplemental Material [30]). Thus, the slight fluctuation in Fig. 3(b) can be attributed to various tip structures (discussed later). Because Eq. (5) states that $t$ is proportional to $F_{FER}^{-\frac{2}{3}}$, $< j >$ on Ag(111) is insensitive to $F_{FER}$ and can be represented by $F_{FER}^{\frac{2}{3}}/\Delta E$. Therefore, according to Eq. (3), the decay rate $D$ on Ag(111) is constant if the tip-structure effect is ignored. The transmission and light emission are two independent pathways for WFD; a constant $D$ indicates that $\Gamma_t$ and $\Gamma_l$ are constant. Because $\Gamma_l$ is irrelevant to the surface orientation and $\Gamma_t$ is 0 for Ag(100), $D$ on Ag(100) should be constant but smaller than that on Ag(111). Although Fig. 3(a) indicates that $D$ on Ag(100) is indeed smaller, the drastic fluctuation of $1/\Delta E$ reflects that $\Gamma_l$ is highly unstable. Figures 3(c) and 3(d) displays that $\Delta E$ can vary by up to a factor of four even under the same $F_{FER}$, which is reminiscent of quantum trapping [24].

Quantum trapping manifesting in $\Delta E$ originates from following mechanisms. (1) Through the exchange interaction, two resonant electrons with opposite spins are successively emitted from the STM tip per unit time. (2) Both electrons leave FER through light emission but one electron emits light first to become a relaxed electron and is momentarily trapped in a potential well beneath the STM tip because of quantum trapping. (3) Both the relaxed electron and resonant electron have the same spin. Because of the Pauli exclusion principle, the resonant electron cannot emit light while the relaxed electron remains trapped in the well. Thus, the lifetime of the resonant electron is controlled by the relaxed electron. Due to these mechanisms, the change of $\Delta E$ is mainly determined by the mean lifetime of resonant electrons subsequently emitting light. On bulk $MoS_2$, the variation in $\Delta E$ can be as high as one order of magnitude whenever the relaxed electrons engage in resonance trapping [24]. Because of the quantum trapping and the action of the Pauli exclusion principle, the WFD of relaxed electrons in the potential well takes over that of resonant electrons emitting light subsequently.

Because a metallic surface cannot be penetrated by an electric field, the potential on the Ag surface remains constant [Fig. 4(a)] even when the tip structure consists of a base with an effective radius of tens of nanometers and a protrusion (marked by an arrow). The open angle of the protrusion defines the sharpness of an STM tip and determines $F_{FER}$. We suggest that a potential well can still be formed if the surface dipole layer is involved. The surface dipole layer originates from the electrons that spill into the vacuum region to form a negatively charge layer [Fig. 4(b)] and thereby positive charge layer is created near the surface [31]. According to electrostatics, the field on a metal surface is constant if the tip base is flat [32]. Let us assume that the constant field is sufficiently strong to block electrons to spill out [Fig. 4(c)]. When the base has curvature, beneath the protrusion, there still exists a local region where the field on the surface remains constant. The field outside this region decreases with the distance from the region center. Consequently, the electron density and its range outside the constant-field region gradually increase from zero to the values in the absence of a field, because of which a potential well for relaxed electrons exists on the surface, as depicted in Fig. 4(d). Figure 4(e) displays that the well size should be smaller than that in Fig. 4(d) when the same protrusion is on a base with a smaller radius. Because quantum trapping is highly sensitive to the well size, Figs. 4(d) and 4(e) reveal why the linewidths in Fig. 3(d) differ considerably even under the same $F_{FER}$ (sharpness). Well sizes are expected to be dissimilar when the tips have the same base but different sharpness levels. Because of the intervention of the quantum trapping sensitive to the tip structure, $\Delta E$ on Ag(100) is unavoidably fluctuated. The narrow FER in Fig. 3(d) is related to resonance trapping. The quantum trapping effect on Ag(100) proves the existence of the surface dipole layer.

Because of the energy gap, light emission is the only channel through which paired resonant electrons can leave FER, enabling the coexistence of a relaxed electron and a resonant electron. If no energy gap exists, resonant electrons can also leave an FER by the transmission. Consequently, the probability that a pair of resonant electrons both emit light is reduced to $\left(\frac{\Gamma_l}{\Gamma_t+\Gamma_l}\right)^2$. Moreover, relaxed electrons engaging in resonance trapping is unable to considerably prolong the lifetime of resonant electrons through the Pauli exclusion principle because of the aforementioned transmission. Because of these two factors, although the surface dipole layer exists, the signal of resonance trapping on Ag(111) is weak, which results in slight fluctuation [Fig. 3(b)].

In summary, we demonstrate that WFD differs noticeably on Ag(111) and Ag(100) surfaces because of their dissimilar band structures. Because Ag(111) has no energy gap, two resonant electrons emitted from the tip through the exchange interaction can leave FER through transmission. Consequently, electrons leaving FER through light emission are fewer than those on Ag(100), which results in an almost stable dissipation rate involving weak quantum trapping signal. By contrast, WFD on Ag(100) is influenced by a strong resonance trapping effect that is related to the structure of the STM tip and the surface dipole layer, causing considerable fluctuation in the dissipation rate. Technically, the drastic fluctuation enables investigation of the surface dipole layer on the metal surface, which cannot be observed through STM under the normal tunneling condition because the electric field in STM junction may alter the electron layer.

The authors are grateful for the support provided by the Ministry of Science and Technology (grant numbers: MOST 109-2112-M-001-049 and MOST 106-2112-M-007-012-MY3) and Academia Sinica (grant numbers: AS-iMATE-109-15), Taiwan. H.T.J. also thanks the CQT-NTHU-MOE, NCHC and CINC-NTU, Taiwan for technical support.

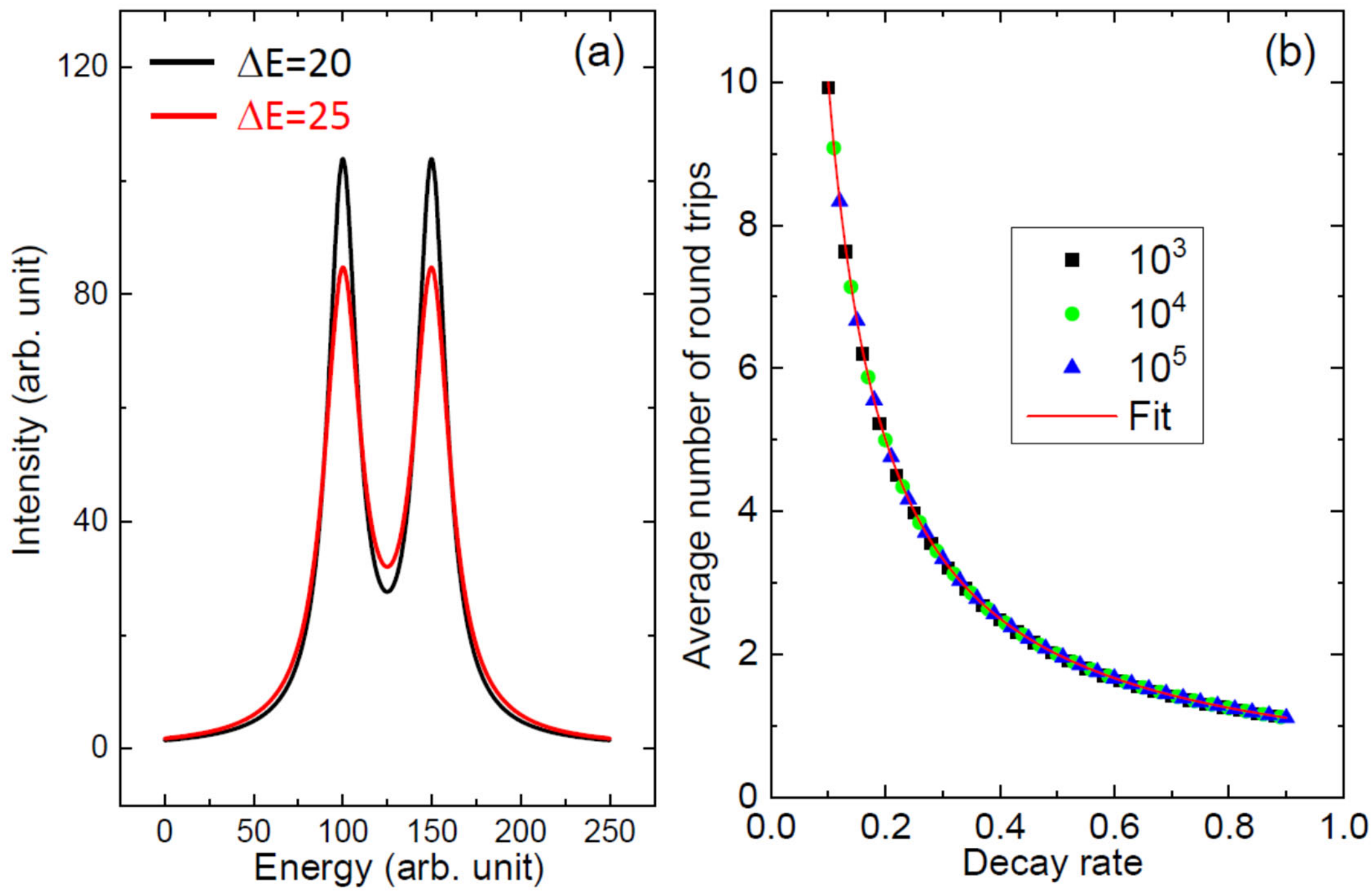

FIG. 1 (a) Under the same intensity integration, the valley intensity between two Lorentzian peaks with linewidth ΔE = 25 is higher than that with ΔE = 20 for a peak separation E = 50. By contrast, the peak intensities for ΔE = 25 are lower than those for ΔE = 20. (b) Average number of round trips $< j >$ versus the decay rate of the probability per round trip $D$ for a number of electrons $N = 10^3$, $10^4$, and $10^5$, showing a decreasing trend that can be fit by a curve representing $< j > = 1/D$.

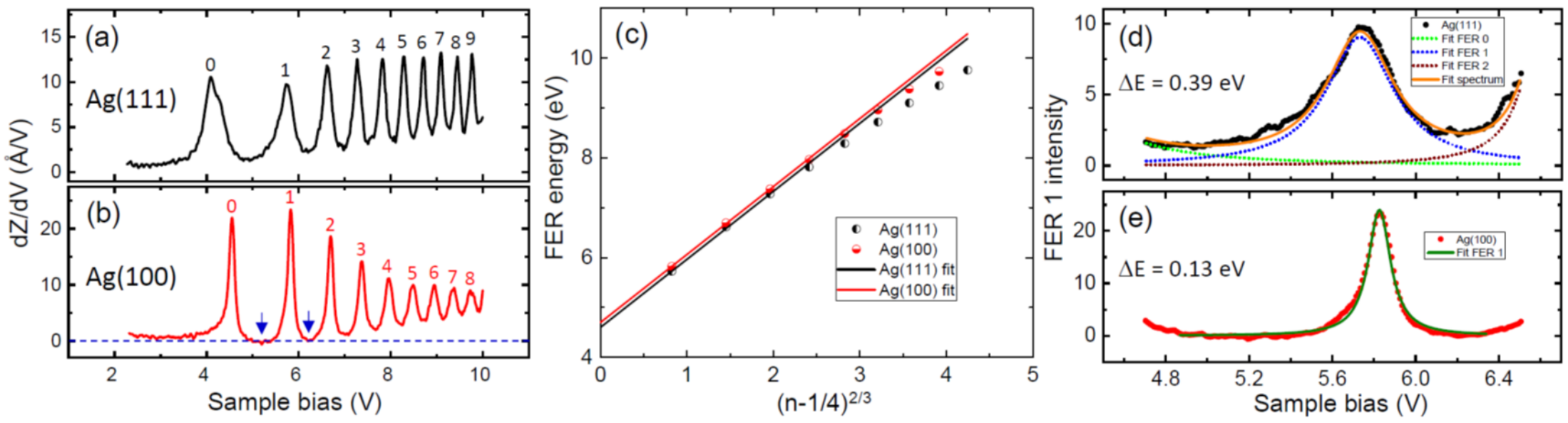

FIG. 2 Typical FER spectra acquired on (a) Ag(111) and (b) Ag(100) in (a) and (b). The numbers mark the orders of FERs. The dashed line in (b) indicates zero spectral intensity. The intensities of the valleys (marked by arrows) around the FER 1 peak are exactly zero, implying that Ag(100) has an energy gap above the vacuum level in its projected bulk band structure and that the energy of FER 1 is in this energy gap. (c) Peak energies of the higher-order FERs in (a) and (b) versus $(n-1/4)^{2/3}$. The data points of two cases are favorably fit by lines for orders from 1 to 3. (d) FER 1 in (a), the ΔE of which is obtained from Lorentzian fitting and decomposition of the intensity superposition with FER 0 and FER 2. (e) FER 1 in (b), ΔE of which is simply obtained from Lorentzian fitting because of zero valley intensity indicating no intensity superposition with FER 0 and FER 2.

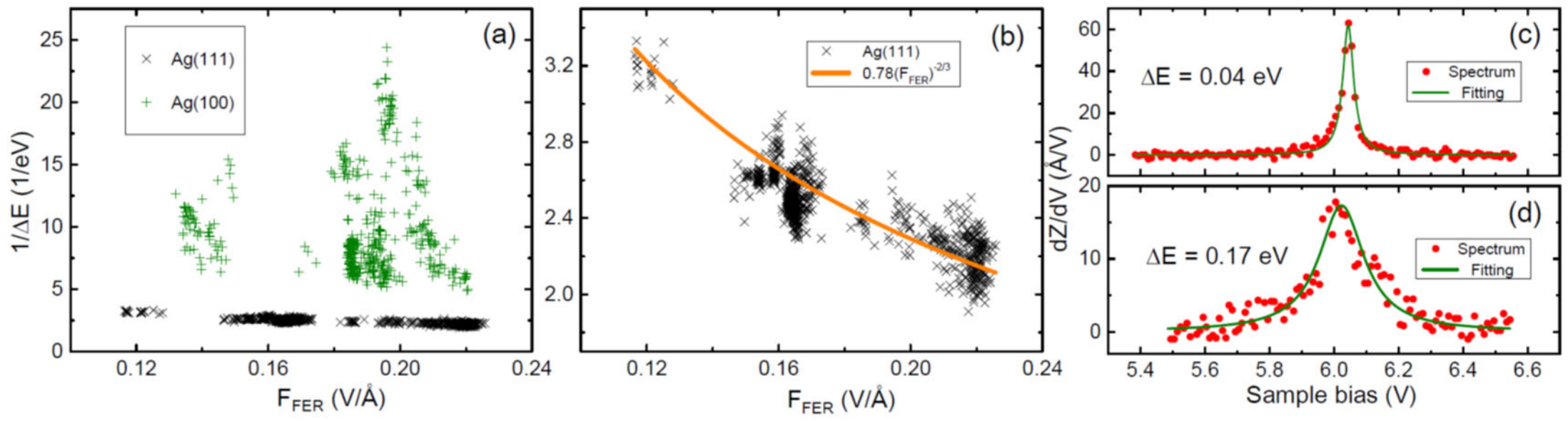

FIG. 3 (a) Plots of $1/\Delta E$ versus $F_{FER}$ on Ag(111) and Ag(100), obtained by accumulating FER spectra of various sharpness levels. (b) $1/\Delta E$ versus $F_{FER}$ on Ag(111) in (a), revealing that on average, data points follow a curve representing $1/\Delta E$ proportional to $F_{FER}^{-\frac{2}{3}}$. (c) FER 1 on Ag(100) with $\Delta E$ = 0.04 eV and (d) FER 1 on Ag(100) with $\Delta E$ = 0.17 eV, acquired under the same $F_{FER}$.

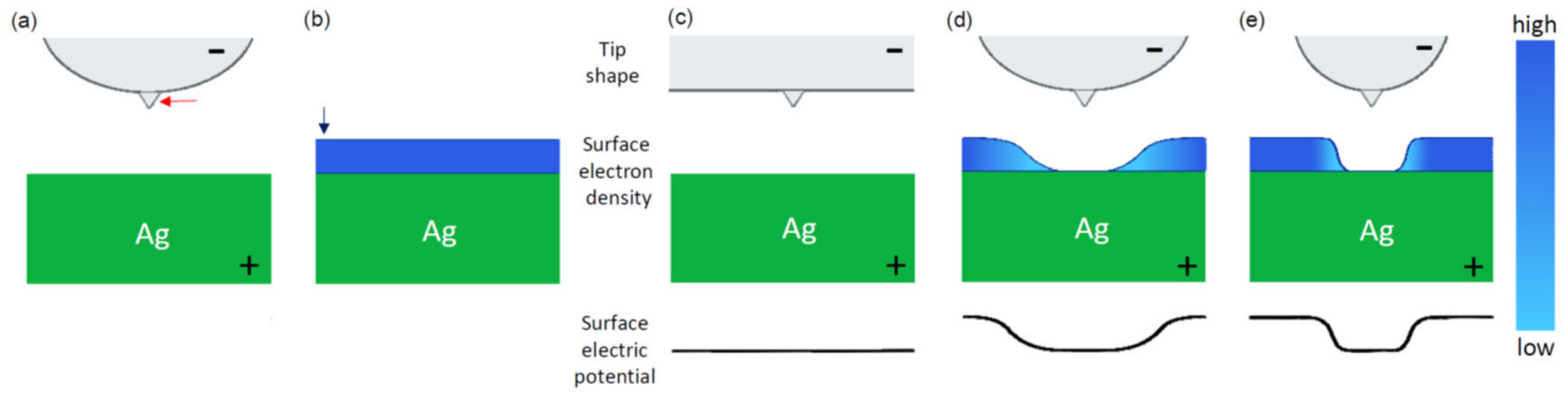

FIG. 4 Step-by-step illustrations of formation of a potential well on the Ag surface beneath an STM tip. (a) The potential on the Ag surface is constant even when the tip consists of a base with an effective radius of tens of nanometers and a protrusion (marked by an arrow). (b) In the free space, the Ag surface naturally has a surface dipole layer, which originates from electrons that spill into the vacuum region to form a negatively charge layer (marked by an arrow). (c) The field on the metal surface is constant if the tip base is flat. The constant field is assumed to be sufficiently strong to prevent electrons from spilling out. The surface electric potential is constant in this situation. (d) When the base has curvature, a local region exists beneath the protrusion in which the field on the surface is constant, and the field outside this region decreases with the distance from the region center. Consequently, the electron density and its range outside the constant-field region gradually increase from zero to the values in the absence of a field, which results in a potential well for electrons on the surface. (e) The well size should be smaller than that in (d) when the same protrusion is on a base with a smaller radius. The scale bar displays the surface electron density.

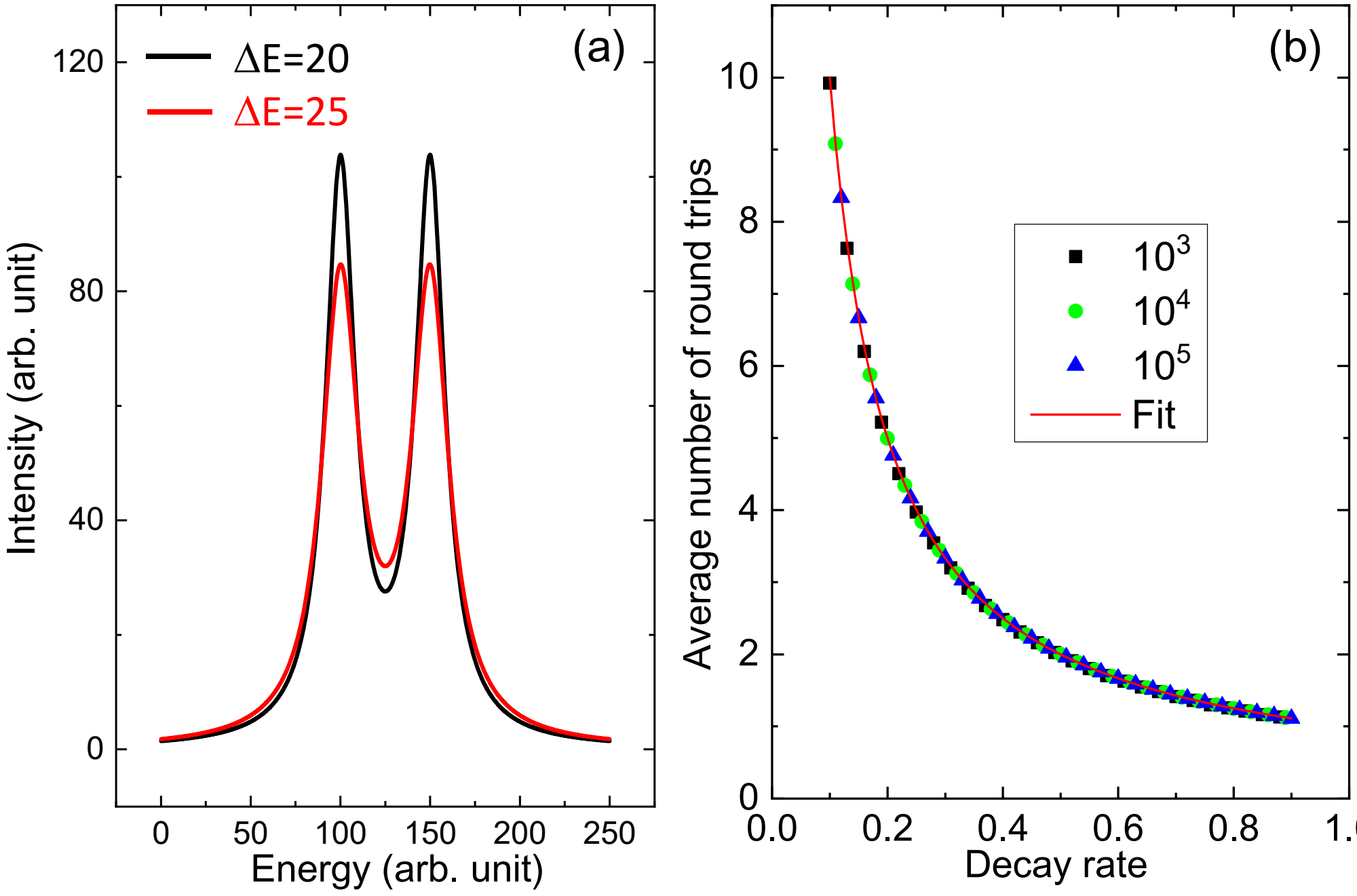

FIG. 1 one column

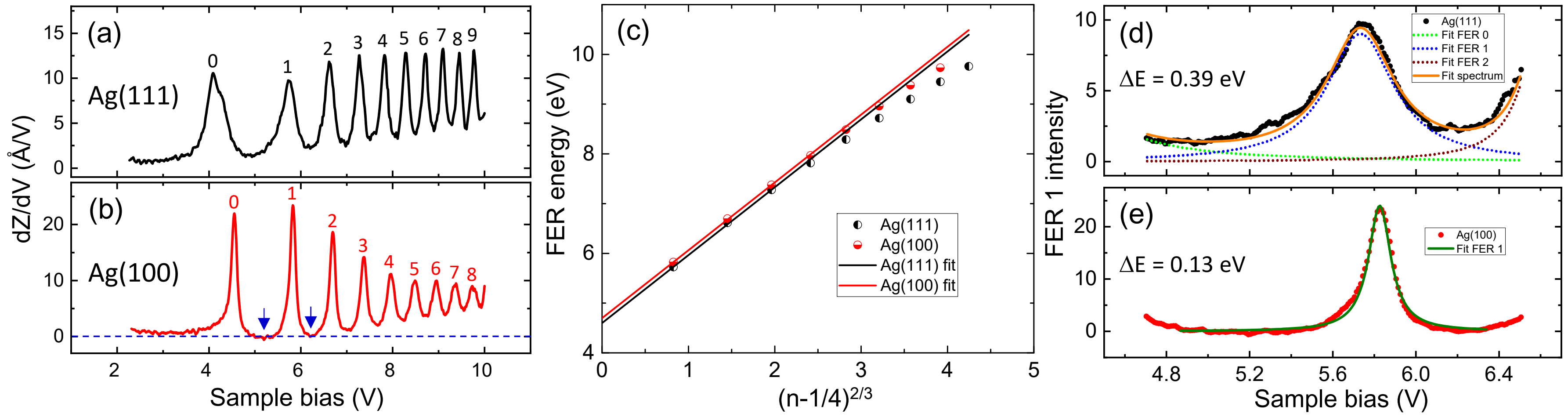

FIG. 2 two columns

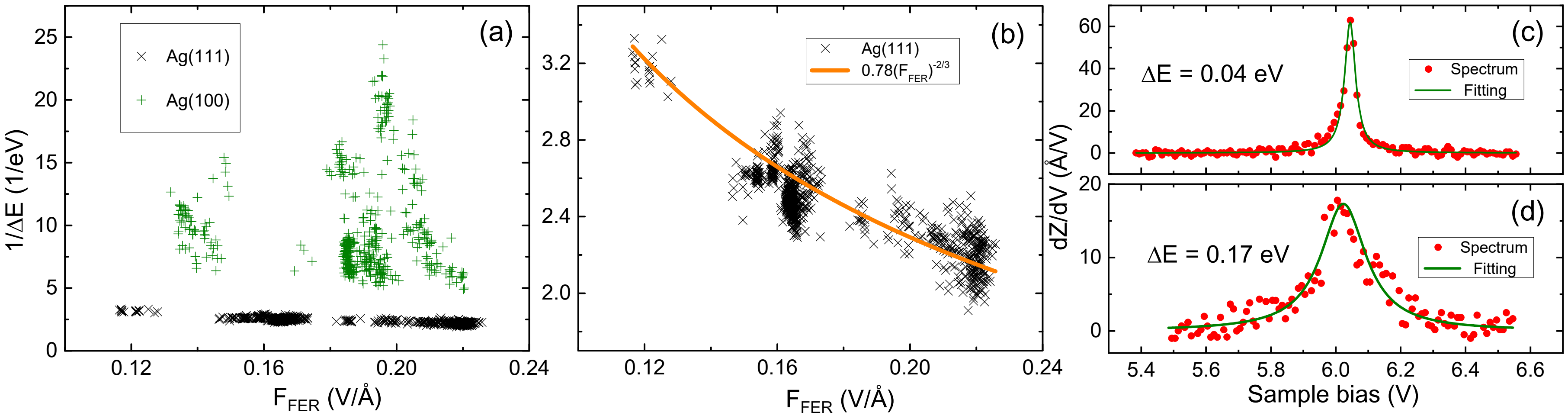

FIG. 3 two columns

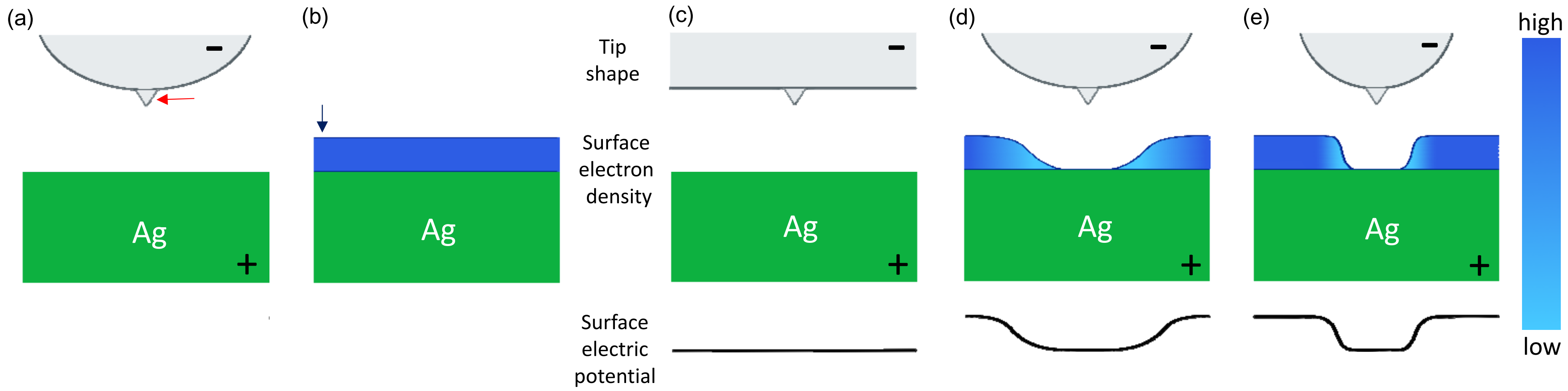

FIG. 4   two columns

## Supplemental Material:

# Impact of Band Structure on Wave function Dissipation in Field Emission Resonance

Wei-Bin Su[1*], Shin-Ming Lu[1], Ho-Hsiang Chang[1], Horng-Tay Jeng[2,1,3*], Wen-Yuan Chan[1], Pei-Cheng Jiang[1], Kung-Hsuan Lin[1], and Chia-Seng Chang[1]

[1]*Institute of Physics, Academia Sinica, Nankang, Taipei 11529, Taiwan*

[2]*Department of Physics, National Tsing Hua University, Hsinchu 30013, Taiwan*

[3]*Physics Division, National Center for Theoretical Sciences, Hsinchu 30013, Taiwan*

[*]Corresponding authors.

wbsu@phys.sinica.edu.tw; jeng@phys.nthu.edu.tw

## 1. Spectra due to spontaneous changes of tip sharpness on Ag(100)

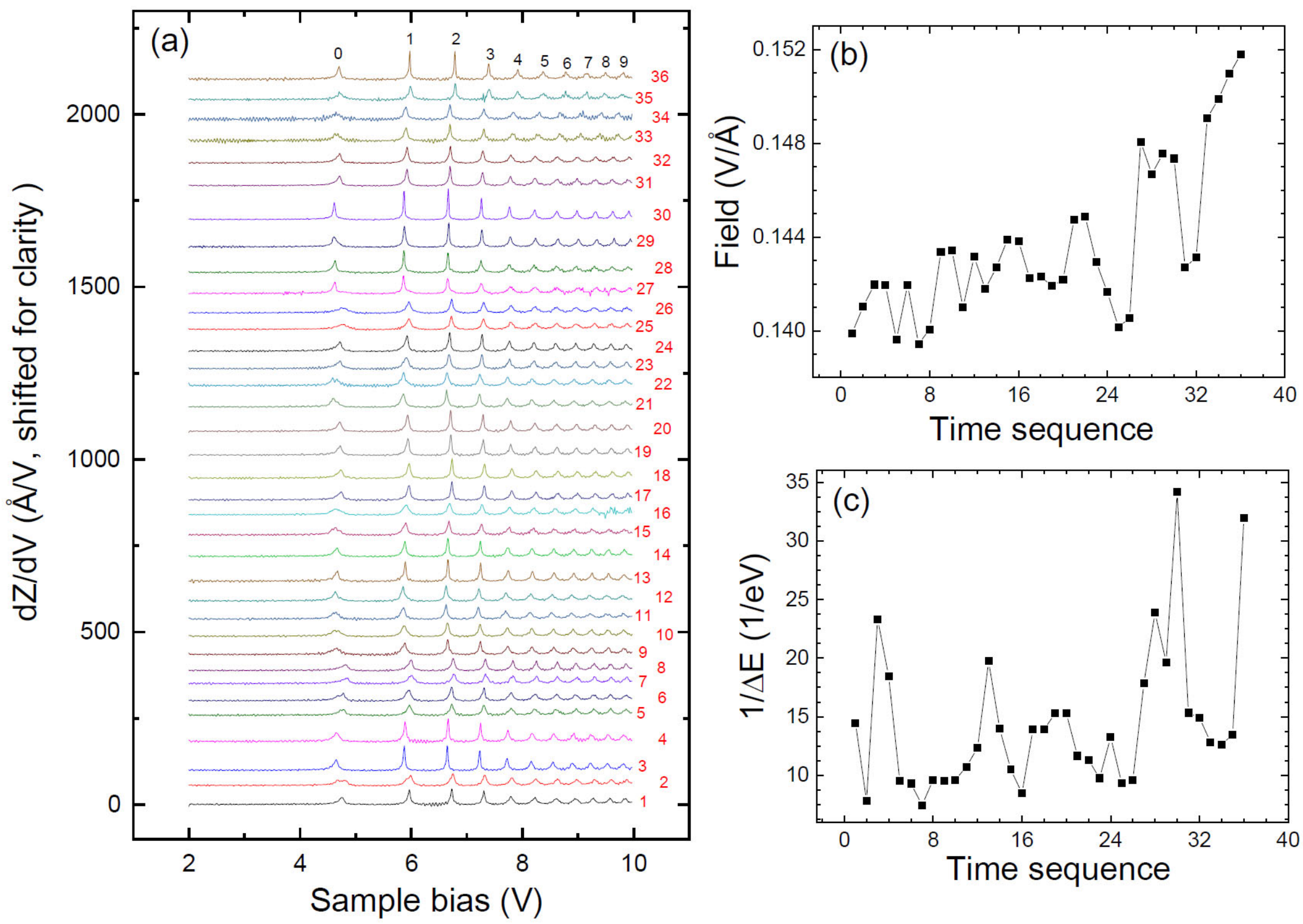

**Figure S1.** (a) Spectra with FERs of different shapes and energies due to spontaneous changes of tip sharpness. The numbers at the right-hand side indicate a time sequence of different spectra. (b) $F_{FER}$ for FERs in each spectrum in (a) versus time sequence, revealing that $F_{FER}$ was fluctuated with time. $F_{FER}$ for FERs in each spectrum was calculated by Eq. (4) in the text. (c) The reciprocal of FER 1 linewidth in each spectrum versus time sequence, also exhibiting fluctuation.

## 2. Spectra of different currents under the same tip structure on Ag(111)

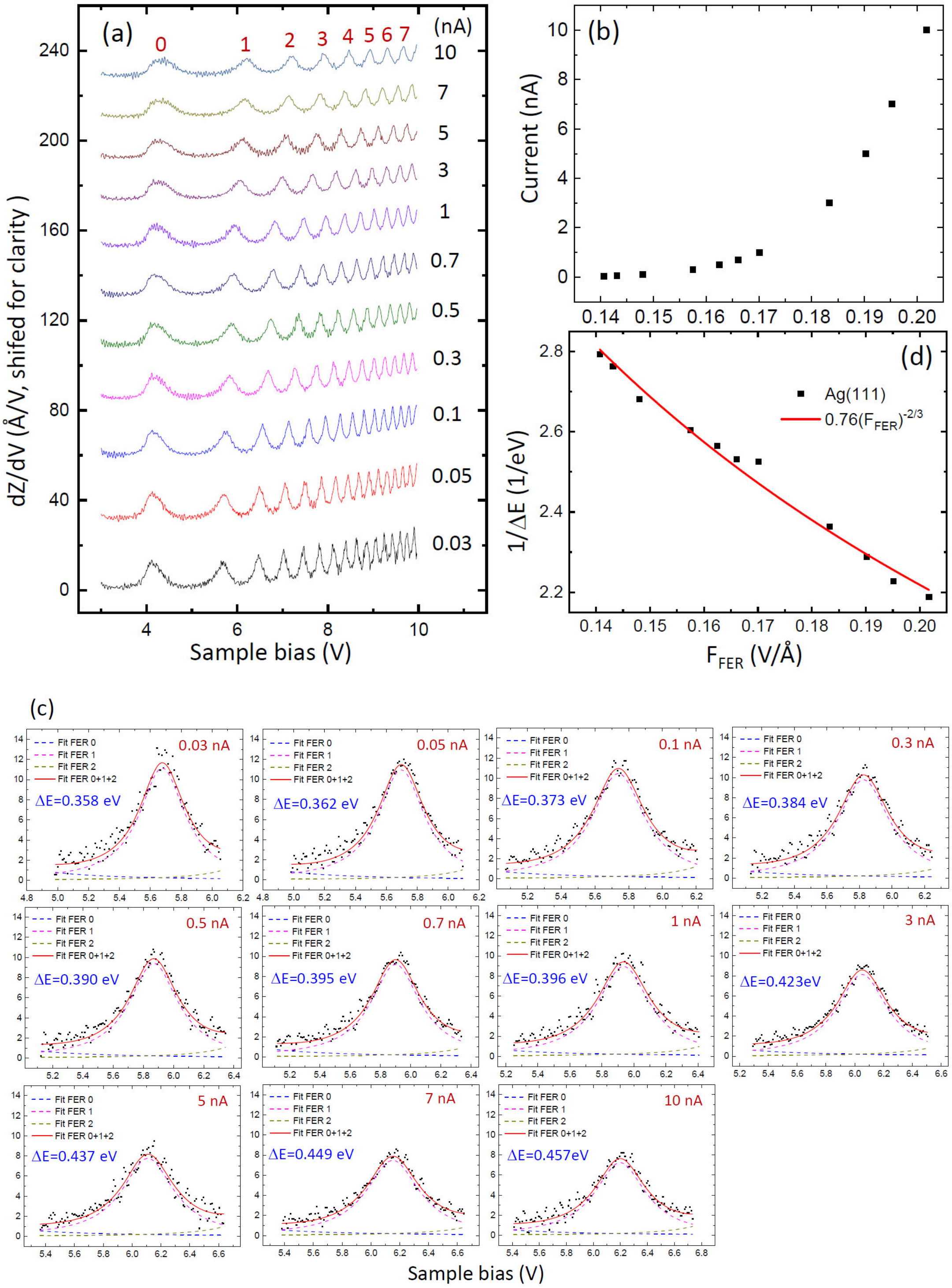

**Figure S2.** (a) FER spectra acquired at 5 K on Ag(111) under various set currents from 0.03 to 10 nA. (b) The set current versus $F_{FER}$ plot displays smooth variation, indicating that the tip structure was unchanged when acquiring FER spectra in (a). (c) Lorentzian fittings of FER 1 in spectra in (a), showing the obtained ΔE of FER 1 increases with increasing currents.